\documentclass[aps,prl,twocolumn,showpacs,superscriptaddress,groupedaddress]{revtex4}  
\usepackage{graphicx}  
\usepackage{dcolumn}   
\usepackage{bm}        
\usepackage{amssymb}   
\usepackage{subfigure}  
\begin{document}

\widetext



\title{Pressure dependence of the ultrasound attenuation and speed in bubbly media: Theory and experiment}

\author{A.J. Sojahrood}

\affiliation{Department of Physics, Ryerson University, Toronto, Ontario, Canada}
\author{Q. Li}

\affiliation{Department of Biomedical Engineering, Boston University, Boston, MA, USA}
\author{H. Haghi}
\affiliation{Department of Physics, Ryerson University, Toronto, Ontario, Canada}

\author{R Karshafian}
\affiliation{Department of Physics, Ryerson University, Toronto, Ontario, Canada}

\author{T.M. Porter}

\affiliation{Department of Biomedical Engineering, Boston University, Boston, MA, USA}

\author{M.C. Kolios}
\affiliation{Department of Physics, Ryerson University, Toronto, Ontario, Canada}

\date{\today}

\begin{abstract}
Results of the measurements of sound speed and attenuation in a bubbly medium are reported. Monodisperse bubble solutions are sonicated with broadband ultrasound pulses with pressure amplitudes ranging between 12.5-100 kPa. Fundamental relationships between the frequency dependent attenuation, sound speed and pressure are established.  A new model for the estimation of sound speed and attenuation is derived that incorporates the effect of nonlinear bubble oscillations on the wave propagation in the bubbly media. Model predictions are in good agreement with experimental results.
\end{abstract}

\maketitle


Acoustically excited microbubbles (MBs) are present in a wide range of phenomena; they have applications in sonochemistry \cite{1}; oceanography and underwater acoustics \cite{2,3}; material science \cite{4}, sonoluminescence \cite{5} and in medicine \cite{6,7,8,9,10,11,12}. Due to their broad and exciting biomedical applications, it has been stated that \H{} The future of medicine is bubbles \H{} \cite{12}. MBs are used in ultrasound molecular imaging \cite{6,7} and recently have been used for the non-invasive imaging of the brain microvasculature \cite{7}. MBs are being investigated for site-specific enhanced drug delivery \cite{8,9,10,11} and for the non-invasive treatment of brain pathologies (by transiently opening the impermeable blood-brain barrier (BBB) to deliver macromolecules \cite{9}; with the first in human clinical BBB opening reported in 2016 \cite{8}). However several factors limit our understanding of MB dynamics which consequently hinder our ability to optimally employ MBs in these applications. The MB dynamics are nonlinear and chaotic \cite{13,14,15}; furthermore, the typical lipid shell coatings add to the complexity of the MBs dynamics due to the nonlinear behavior of the shell (e.g., buckling and rupture \cite{16}). Importantly, the presence of MBs changes the sound speed and attenuation of the medium \cite{17,18,19,20,21}. These changes are highly nonlinear and depend on the MB nonlinear oscillations which in turn depend on the ultrasound pressure and frequency, MB size and shell characteristics \cite{17,18,19,20}.
The increased attenuation due to the presence of MBs in the beam path may limit the pressure at the target location. This phenomenon is called pre-focal shielding (shadowing) \cite{21,22}. Additionally, changes in the sound speed can change the position and dimensions of the focal region; thus, reducing the accuracy of focal placement (e.g., for targeted drug delivery). In imaging applications, MBs can limit imaging in depth due to the shadowing caused by prefocal MBs \cite{21,22}. In sonochemistry, changes in the attenuation and sound speed impact the pressure distribution inside the reactors and reduces the procedure efficacy \cite{23}.\\
An accurate estimation of the pressure dependent attenuation and sound speed in bubbly media remains one of the unsolved problems in acoustics \cite{24}. Most current models are based on linear approximations which are only valid for small amplitude MB oscillations \cite{17}. Linear approximations, however, are not valid for the typical exposure conditions encountered in biomedical applications. In an effort to incorporate the nonlinear MB oscillations in the attenuation estimation of bubbly media, a pressure-dependent MB scattering cross-section has been introduced \cite{2,25}. While the models introduce a degree of pressure dependency (e.g. only the pressure dependance of the scattering cross section were considered while the damping factors were estimated using the linear model), they still incorporate linear approximations for the calculation of the rest of the  damping factors (e.g. liquid viscous damping, shell viscous damping and thermal damping). Additionally, they neglect the nonlinear changes of the sound speed in their approximations. Louisnard \cite{18} and Holt and Roy \cite{26} have derived models based on employing the energy conservation principle. In Louisnard\'{}s approach \cite{18} the pressure dependent imaginary part of the wave number is calculated by computing the total nonlinear energy loss during bubble oscillations. However, this method still uses the linear approximations to calculate the real part of the wave number; thus, it is unable to predict the changes of the sound speed with pressure. Holt and Roy calculated the energy loss due to MB nonlinear oscillations and then calculated the attenuation by determining the extinction cross-section. Both approaches in \cite{18} and \cite{26} use the analytical form of the energy dissipation terms.  In the case of coated MBs with nonlinear shell behavior, such calculations are complex and can result in inaccuracies. The existing approaches for sound speed computations based on the Woods model \cite{26,27} are either limited to bubbles whose expansion are essentially in phase with the rarefaction phase of the local acoustic pressure, or encounter difficulties in nonlinear regimes of oscillations because of their dependence on $\frac{dP}{dV}$  (e.g. \cite{28}) where P is pressure and V is the MB volume.\\
Experimental investigation of the pressure and frequency dependence of the attenuation of bubbly media has been limited to few studies of coated MBs suspensions \cite{29,25}. Importantly, the pressure dependence of the sound speed in bubbly media has not been experimentally investigated. In the absence of a comprehensive and reliable nonlinear model to calculate the sound speed and attenuation, the relationship between the changes in the pressure and variations in the sound speed and attenuation has remained incomplete.\\ The objective of this work is to provide a new model  describing the relationship between the acoustic pressure and the sound speed and attenuation in a bubbly medium, as well as test the predictions experimentally.\\ Here, we report on our controlled observations on the pressure dependence of the sound speed of a bubbly medium. A theoretical model is derived that can predict the pressure dependent attenuation and sound speed; the model is free from any linear approximations and treats the MB oscillations with their full nonlinearity. We have first proposed this model in \cite{30,31}, and reported its initial experimental validation in \cite{32}; initial results were featured as conference papers in \cite{31,33}.\\
 To drive the model, we start with the Caflisch equation \cite{34} for the propagation of the acoustic
waves in a bubbly medium:
\begin{equation}
{\nabla{}}^2\left(P\right)=\frac{1}{C_l^2}\frac{{\partial{}}^2P}{\partial{}t^2}-{\sum_{i=1}^N\
\ \rho_l{}}\frac{{\partial{}}^2{\beta{}}_i}{\partial{}t^2}\ \ \ \
\left(1\right).
\end{equation}
In this equation, P is pressure,$\ C_l$ is the speed of sound in the liquid in the absence of bubbles,
${\rho{}}_l$ is the liquid density and  ${\beta{}}_i$ is the local volume fraction occupied by the gas at time $t$ of  the  \textit{ith} MBs. ${\beta{}}_i$is given by
${\beta{}}_i(t)=\frac{4}{3}\pi{}{R_i(t)}^3N_{i}$  where $R_i$\textit{ (t)} is the
instantaneous radius of the MBs with initial radius of $R{_{0i}}$ and $N_{i}$ is the number of the corresponding MBs per unit volume in the
medium and summation is performed over the whole population of the MBs.   Eq. 1 can also be written in terms of the complex conjugate of the
acoustic pressure. To calculate the attenuation and sound speed we need to determine the wave
number (\textit{k=k$_{r}$-i$\alpha{}$}); the sound speed can be calculated from
\textit{k$_{r }$} which is the real part of the wave number, and the
attenuation \textit{$\alpha{}$} from the imaginary part of the wave number. To
obtain the expressions for the imaginary and real part of the wave number, we
first need to write Eq. 1 and its complex conjugate in the form of the Helmholtz equation; then the
nonlinear wave number will be given by: $k^2=-\frac{{\nabla{}}^2\left(P\right)}{P}\ \ \ \
$.\\ To achieve this, Eq. 1 was multiplied by  $\frac{\ \bar{P}}{P\bar{P}}\ $
and  its complex conjugate was multiplied by   $\frac{P}{P\bar{P}}$ where $\bar{P}$ is the complex conjugate of $P$.  The pressure dependent real and
imaginary parts of $k^2$ were derived using the time average of the results of the
addition and subtraction of the new equations and applying the boundary
conditions of the problem:
\begin{equation}
\langle\Re(k^2)\rangle=\frac{-{\omega{}}^2}{C_l^2}-\frac{2{\rho_l{}}}{T{\left\vert{}P\right\vert{}}^2}\sum_{i=1}^N\int_0^T{\Re(P)}\frac{{\partial{}}^2{\beta{}}_i}{\partial{}t^2}dt
\end{equation}
\begin{equation}
\langle\Im(k^2)\rangle =-\frac{2{\rho_l{}}}{T{\left\vert{}P\right\vert{}}^2}\sum_{i=1}^N\int_0^T {\Im(P)}\frac{{\partial{}}^2{\beta{}}_i}{\partial{}t^2}dt
\end{equation}
where $\Re$ and $\Im$ respectively denote the real and imaginary parts,
\textit{$<$$>$} denotes the time average, $\omega{}$ is the angular frequency of a
propagating wave, and \textit{T} is the  time averaging interval. The
contribution of each MBs with ${\beta{}}_i$ is summed. Using Eqs.
2 and 3, we can now calculate the pressure-dependent sound speed and
attenuation in a bubbly medium. To do this, the radial oscillations of the MBs in
response to an acoustic wave were calculated first. Then equation 2 and 3 were
solved by integrating over the  ${\beta{}}_i$ of each of the MBs in the
population.\\To verify the model in the linear oscillation regimes, predictions from the
model were compared to the results of published models simulating linear forced
oscillations of uncoated bubbles (Commander and Prosperetti \cite{17}).

\begin{figure*}
\includegraphics[scale=0.5]{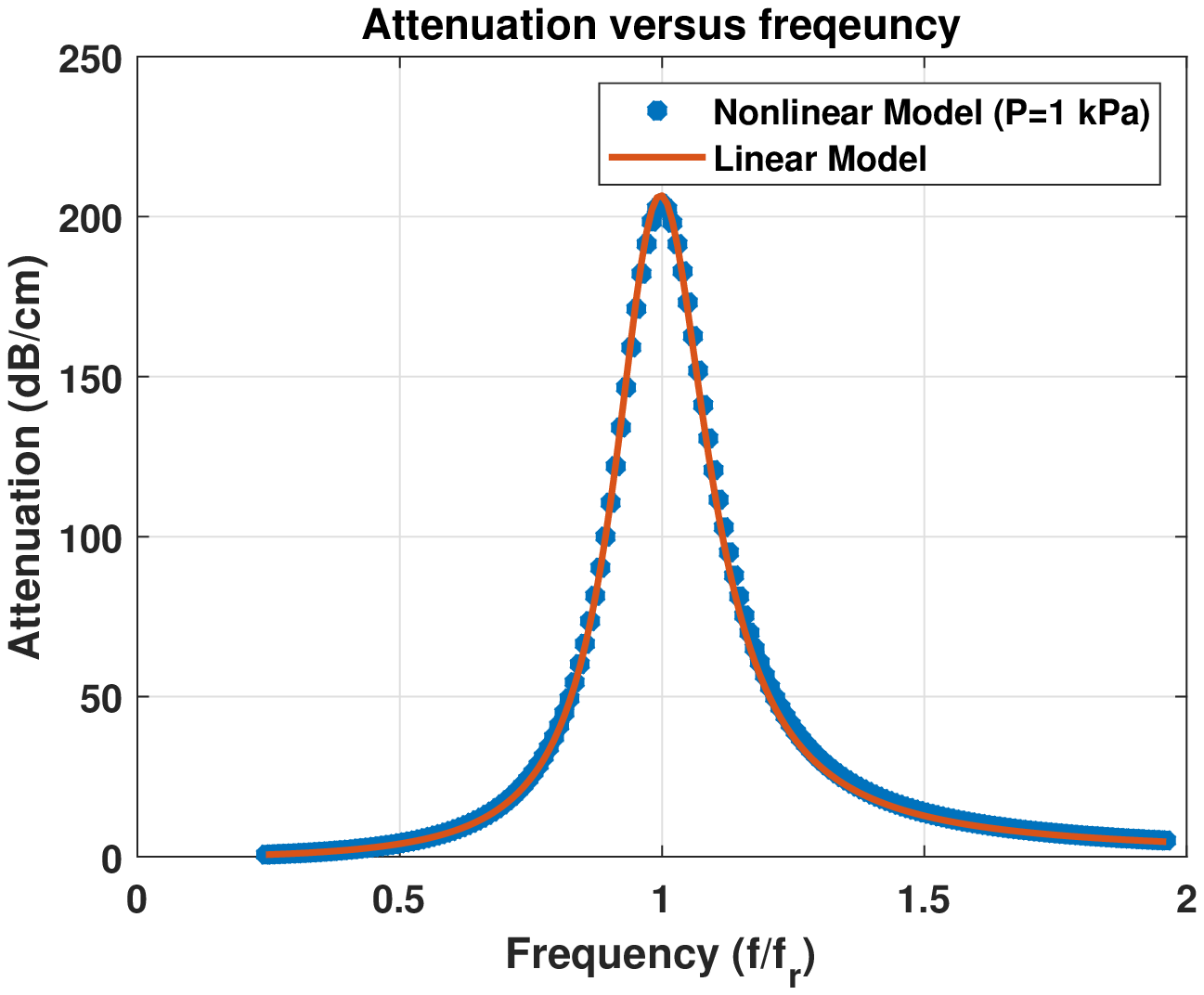} \includegraphics[scale=0.5]{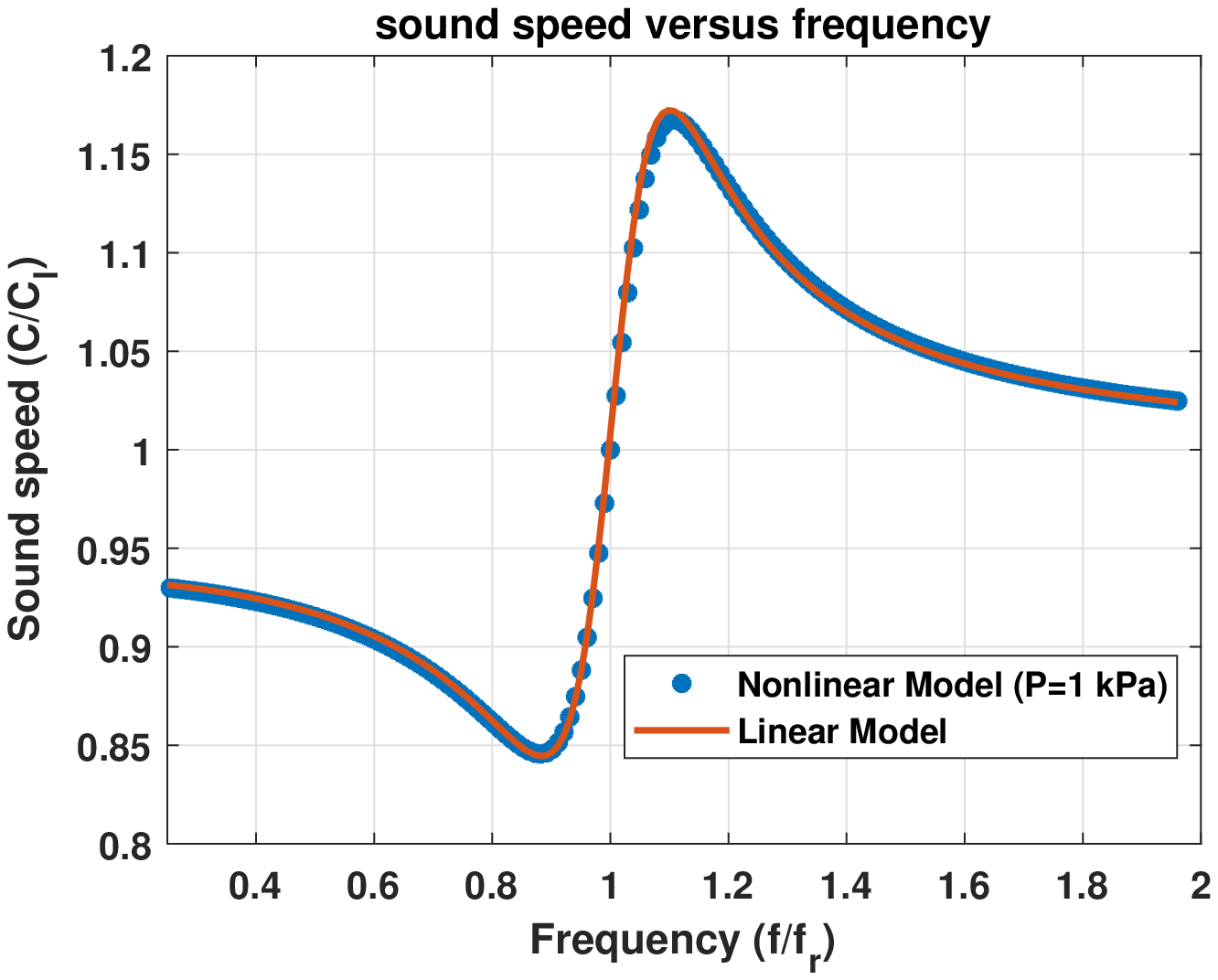} \\
(a) \hspace{6cm} (b)\\
\includegraphics[scale=0.5]{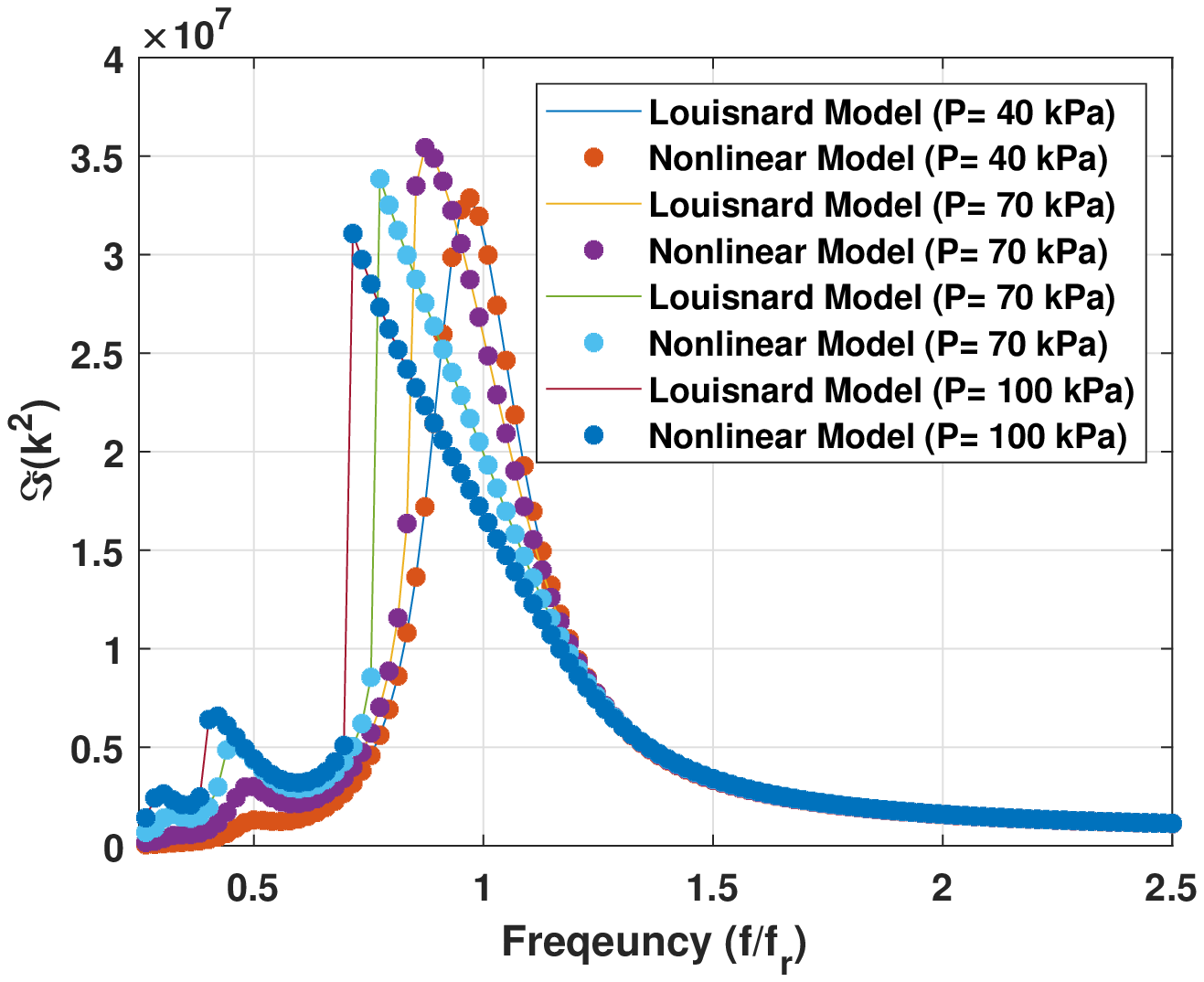}\includegraphics[scale=0.5]{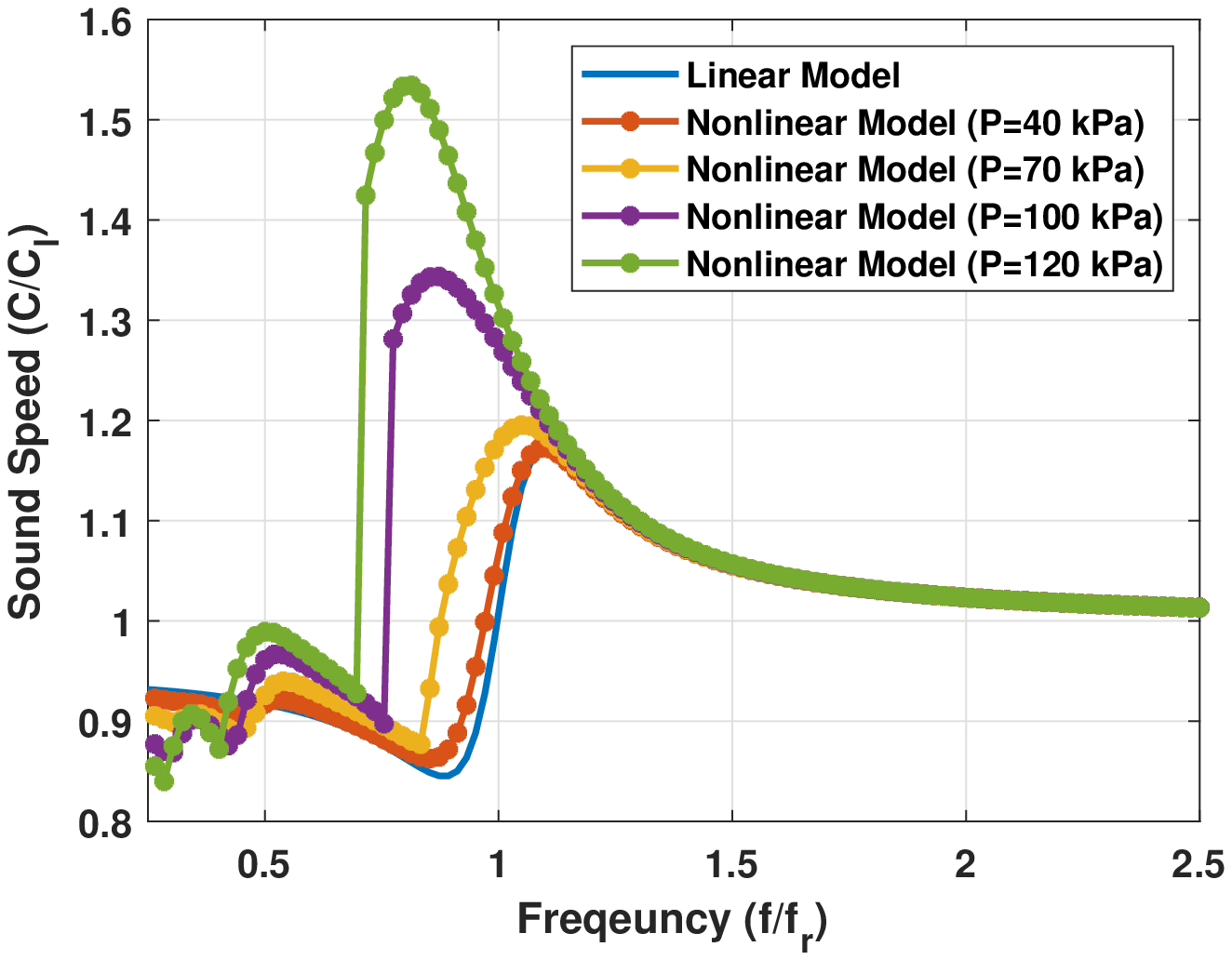}\\
(c) \hspace{6cm} (d)
\caption{ Case of a bubbly medium with uncoated MBs with $R_0= 2 \mu m$ and $\beta=10^{-5}$ a)  Attenuation calculated using the linear model and nonlinear model at P= 1kPa,  b) sound speed calculated using the linear model and the nonlinear model at (P= 1kPa) and c)  Pressure dependent time-averaged $\Im$ part of $k^2$ as calculated by Louisnard (linear) and nonlinear model   d) Pressure dependent sound speed as calculated by the Louisnard and the nonlinear model.}
\end{figure*}

\textbf{Linear regime of oscillations:} Keller-Miksis model for large MB oscillations \cite{19,20} was coupled with the ordinary differential equations
describing the thermal damping effect \cite{19,35}. The new set of differential
equations were solved to calculate the MB radial oscillations.
Constants of the linear model were derived using the approach given by \cite{17}.
Since the linear model is only valid for narrowband pulses with small pressure
amplitudes, pulses of \textit{1kPa} amplitude with 60 cycles were chosen at
each frequency, and the last 20 cycles were used for the integration using Eqs. 2
and 3. For the linearized model, the initial MB diameter is 4 $\mathrm{\mu}$m;
the gas inside the MB is\textit{ air} and the thermal properties are chosen
from \cite{36} and $\beta{}$ was set to 10$^{-5}$. Figures 1a-b compare the
attenuation and sound speed predictions between the linear model and the
non-linear model given by Eq. 2 and 3. Model predictions are in excellent
agreement with the linear model for small amplitude radial oscillations.\\ \textbf{Non-Linear regime of oscillations:}As the pressure increases, assumptions (e.g. small amplitude MB oscillations) on which the linear model is based on are no longer valid. To investigate
the effect of pressure, the radial oscillations of the MBs were simulated for
exposures of acoustic pressure amplitudes of 40, 70, 100 and 120 kPa. Using the
approach in \cite{18,19}, we derived the energy expressions for nonlinear damping due
to liquid viscosity, radiation, and thermal contributions. The
imaginary and real part of the wave number were then calculated using Eqs. 2 and 3
and the expressions that are given by the Louisnard model \cite{18}. The predictions of the two
models are illustrated in Figure 1c-d. Figure 1c shows that the
$\left\langle{}{\Im}\left(k^2\right)\right\rangle{}$ calculated by Eq. 3 is in
excellent agreement with the Louisnard model for all the acoustic pressures that
are investigated here. However, the Louisnard model uses the linear model to
calculate the $\left\langle{}{\Re}\left(k^2\right)\right\rangle{}$. Figure 1d, shows
that sound speed changes with pressure, and predictions of Eq. 2 significantly deviate from the linear values. Our model incorporates the pressure-dependent changes in $\left\langle{}{\Re}\left(k^2\right)\right\rangle{}$ and
thus can be used to predict the changes of the sound speed with pressure. To our best knowledge this is the first time that the frequency-pressure dependence of the sound speed in a bubbly medium has been characterized.\\
As the pressure increases, the resonance frequency of the bubbles decreases \cite{36}, which is observed as the peak of $\Im{(k^2)}$ shifts towards lower frequencies; this corresponds to the frequencies at which the sound speed in the bubbly medium is equal to the sound speed in the absence of the bubbles. At these frequencies and for pressure dependent resonances, the oscillations are in phase with the driving acoustic pressure.  As the pressure increases, the maximum sound speed of the bubbly medium increases and the peak of sound speed occurs at a lower frequency, which depends on the driving acoustic pressure amplitude \cite{36}. The abrupt increase in the sound speed at particular frequencies in Fig 1d is due to the pressure dependent resonance frequency which is described in detail in \cite{36}. We have previously shown that when MBs are sonicated with their pressure dependent resonance frequency, radial oscillations amplitude of the MBs undergo a saddle node bifurcation (rapid increase in amplitude)as soon as the pressure increases above threshold\cite{36}.\\
To experimentally explore  the pressure-dependent changes of the sound speed and
attenuation for coated MBs, monodisperse lipid shell MBs were produced
using flow-focusing in a microfluidic device as previously described \cite{29,37}.
Figure 2 shows the size distribution of the MBs in our experiments. The setup
for the attenuation and sound speed measurements is the same as the one used in \cite{37}. The thermal properties for the gas can be found in \cite{38}. A pair of
single-element 2.25-MHz unfocused transducers (Olympus, Center Valley,~PA;
bandwidth 1-3.0 MHz) were aligned coaxially in a tank of deionized water and
oriented facing each other. Monodisperse MBs were injected into a sample chamber
that was made with a plastic frame covered with an acoustically transparent thin
film. The dimensions of sample chamber were 1.4 x 3.5 x 3.5 cm (1.4 cm acoustic
path length), and a stir bar was used to keep the MBs dispersed. The
transmit transducer was excited with a pulse generated by a pulser/receiver
(5072PR, Panametrics, Waltham,~MA) at a pulse repetition frequency (PRF) of 100
Hz. An attenuator controlled the pressure output of the transmit transducer
(50BR-008, JFW, Indianapolis, IN), which was calibrated with a 0.2-mm broadband
needle hydrophone (Precision Acoustics, Dorset, UK). Electric signals generated
by pulses acquired by the receive transducer were sent to the Gagescope
(Lockport, IL) and digitized at a sampling frequency of 50MHz. All received
signals were recorded on a desktop computer (Dell, Round Rock, TX) and processed
using Matlab software (The MathWorks, Natick, MA).~ The peak negative pressures
of the acoustic pulses that are used in experiments were 12.5, 25 and 50 and 100
kPa.

\begin{figure}
\includegraphics[scale=0.5]{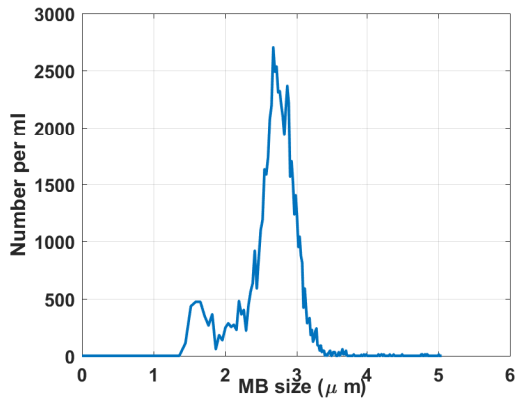}
\caption{ Size distribution of the MBs in the experiments}
\end{figure}

\begin{figure*}
\includegraphics[scale=0.4]{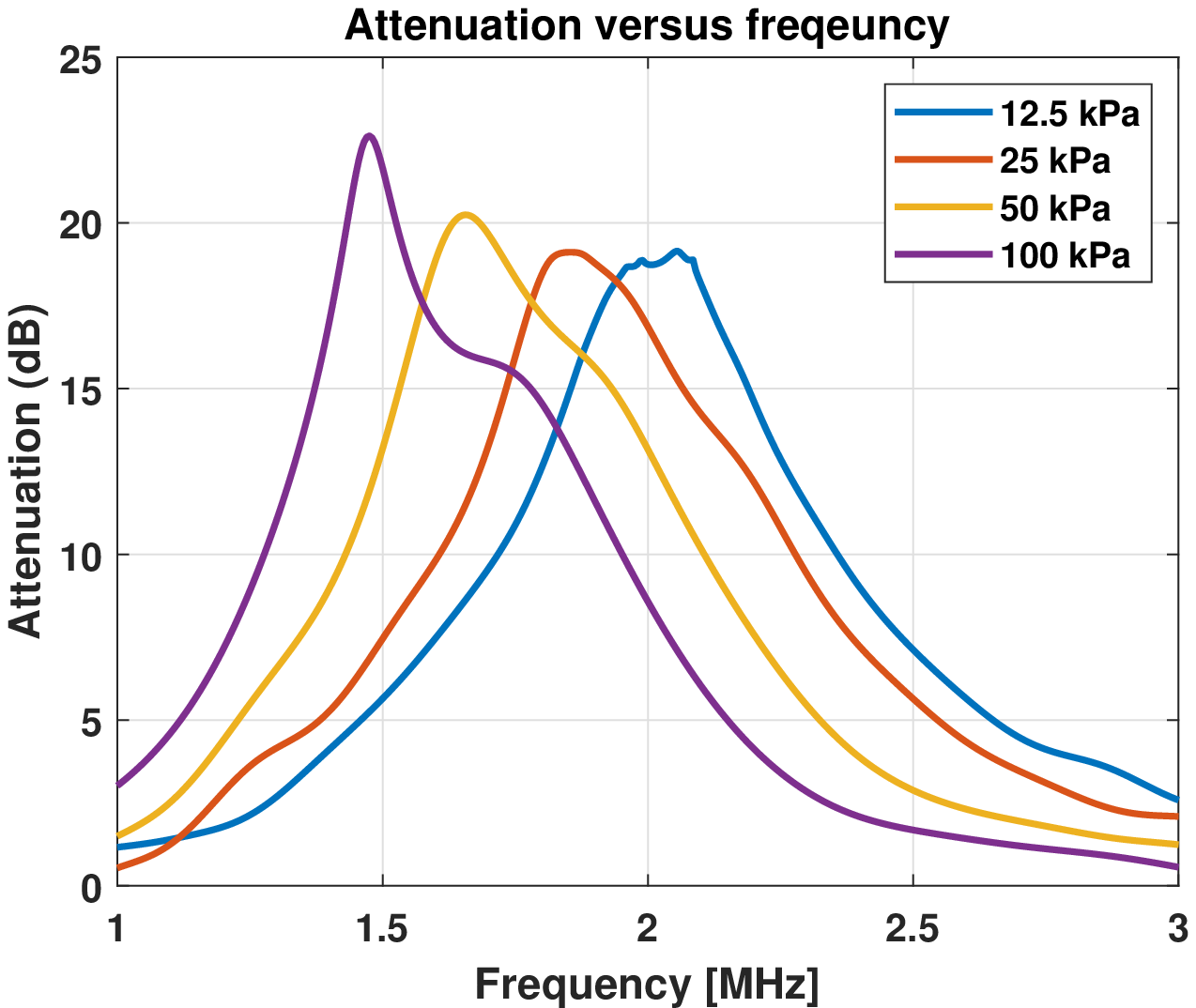} \hspace{1cm} \includegraphics[scale=0.4]{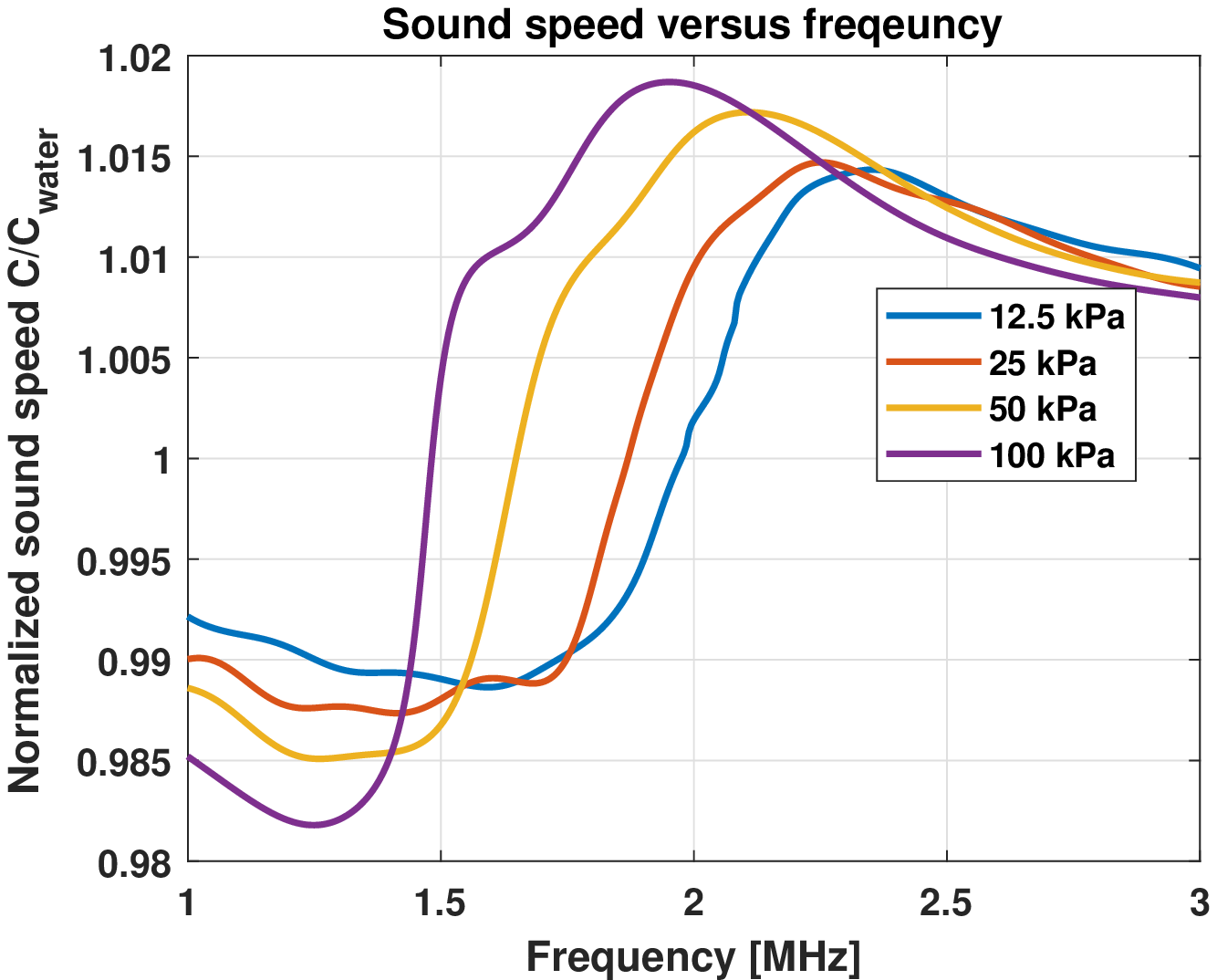}
\caption{ Experimentally measured a) attenuation  and b) sound speed of the bubbly medium}
\end{figure*}

To numerically simulate the attenuation and sound speed, the
Marmottant model \cite{16}, which accounts for
radial-dependent shell properties, was modified to include MB multiple
scattering using the approach introduced in \cite{39}. At each frequency and pressure,
63 MBs were selected from the size distribution in Fig. 2 and were randomly
distributed in a cube with sides of 0.1 mm in length (to simulate the concentration of 63000 MBs/ml in experiments). The
results were weighted by the number density of the MBs in Fig. 2 and were
implemented in Eqs. 2 and 3. This procedure was repeated by iterating different
values for the shell elasticity (0.1N/m$<$\textit{$\chi{}$} $<$2.5 N/m), initial
surface tension (0 N/m $<$ \textit{$\sigma{}$}($R_0)\ $$<$ 0.072 N/m),  shell
viscosity (1e-9 \textit{Pa.s} $<$${\mu{}}_s$$<$$\ 1e-7\ Pa.s)$, and break up
radius $(R_0<R_{breakup}$ $<$ 2$R_0)\ $to fit the shell parameters.\\ Figures 3a-b show the experimentally measured attenuation and sound speed of the mixture respectively. The attenuation of the bubbly medium increases as the pressure increases from 12.5 kPa to 100 kPa and the frequency of the maximum attenuation decreases from ~2.045 MHz to ~1.475 MHz. Maximum sound speed of the medium increases with pressure and the corresponding frequency of the maximum sound speed decreases by pressure increases. To our best knowledge this is the first experimental observation of the pressure dependance of the sound speed.\\
To compare the predictions of the model with experiments, figures 4a-h illustrate the results of the experimentally measured (blue) (with standard deviations of the 100 data point at each condition) and
numerically simulated (red) sound speed of the medium as a function of frequency,
for 4 different pressure exposures of 12.5, 25, 50 and 100 kPa.
The shell parameters that were used to fit the experimental results are \textit{$\chi{}$}=1.1 N/m,
\textit{$\sigma{}$}($R_0)$= 0.016N/m$,\ {\mu{}}_s$=7e-9 $Pa.s$ and
$R_{breakup}$=1.1$R_0$.\\
Figures 4 a-d show the results of the experimentally measured
(blue) and numerically simulated (red) medium attenuation for peak acoustic
pressures of 12.5, 25, 50 and 100 kPa. As the pressure increases, the frequency
at which the maximum attenuation occurs (which indicates the resonance frequency)
decreases (from ~2.02 MHz at 12.5 kPa to 1.475 MHz at 100 kPa) and the magnitude of
the attenuation peak increases (from 16.5 dB/cm at 2 MHz to 21.8 dB/cm at 1.475 MHz).
At 12.5 kPa and for frequencies below $\sim$ 2
MHz, the speed of sound in the bubbly medium is smaller than the sound speed of
water. Above 2 MHz, speed of sound increases and reaches a maximum at 2.25 MHz
with a magnitude of $\sim$ 1.015$C_l$.  At $\sim$
2MHz the sound speed is equal to $C_l$. This is also the frequency where the attenuation peak is maximum in fig 2a. According to the linear theory
\cite{35} at resonance frequency the sound speed of the bubbly medium is equal to the sound speed of the medium without the bubbles. As the pressure increases to 25, 50 and 100 kPa
 the frequency at which the speed of sound in the bubbly sample is
equal to $C_l$ decreases to 1.87, 1.65 and 1.48 MHz respectively. The
frequency at which the maximum sound speed occurs decreases as the pressure increases and the magnitude of
the maximum sound speed increases to $\sim$ 1.019$C_l$
at 100 kPa. The minimum sound speed decreases from $\sim$ 0.989$C_l$ at 12.5 kPa and (1.6 MHz) to $\sim$ 0.981$C_l$ at 100 kPa
and (1.25 MHz).
\begin{figure*}
\includegraphics[scale=0.17]{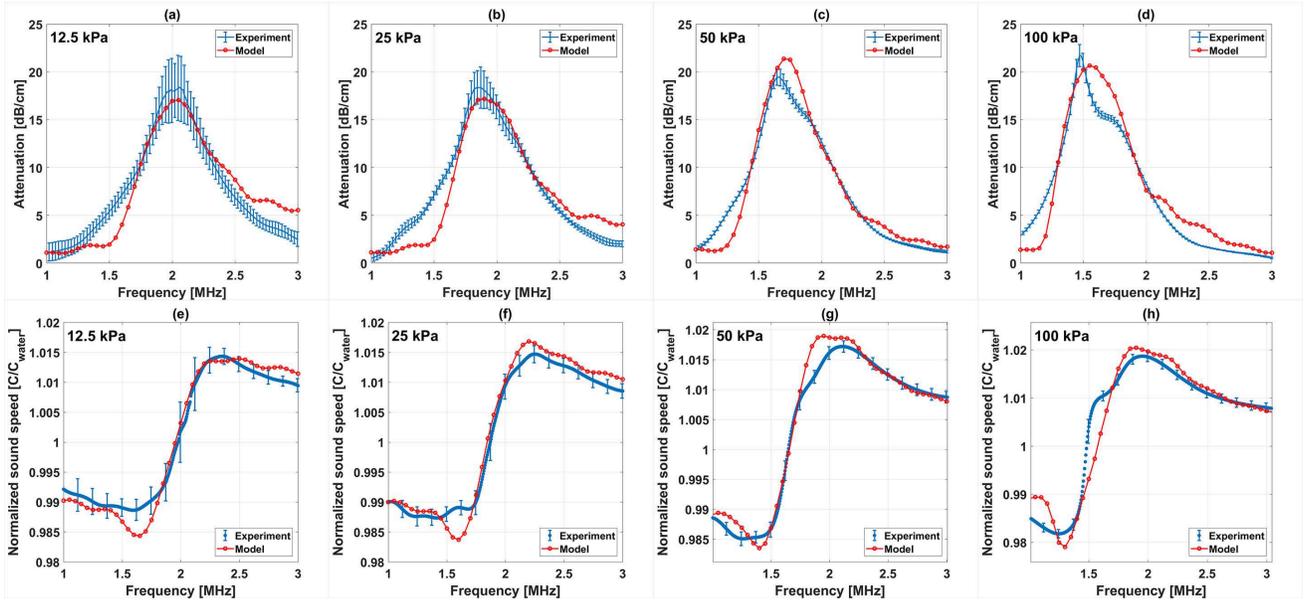}
\caption{ Experimentally measured (blue) and simulated (red) attenuation of the
sample for a) 12.5 kPa, b) 25 kPa, c) 50 kPa and d) 100 kPa . Sound speed of the
sample for e) 12.5 kPa, f) 25 kPa, g) 50 kPa and h) 100 kPa . Errors bars represent the standard deviation.}
\end{figure*}
At each pressure, the frequency at which attenuation is maximized (pressure dependent resonance) is approximately equal to the frequency where the sound
speed becomes equal to$C_l$. Thus, it can be postulated that, even at
the pressure dependent resonance frequency (a nonlinear effect) \cite{14,36}, the phase
velocity is in phase with the driving force \cite{40}. This observation is also consistent with the numerical results of the uncoated bubble in figure 1c-d.
The numerical results are in good agreement with the experimental observations.\\
The small discrepancies can be due to the fact that we
have not considered possible pressure variations within the MB chamber. Additionally we assumed that all the MBs have the identical
shell composition and properties, and effects like strain softening or shear thinning of the shell \cite{41}, and possible MB destruction, were neglected.
As the purpose of the current work was to investigate the pressure dependence of the sound speed and attenuation and to develop
a model to accurately predict these effects, we have the simplest model for lipid coated MBs. Investigation of models that incorporate more complex rheological behaviors of shell
which takes into account the effect of shell properties on sound speed and attenuation is the subject of future works.\\
The changes of the magnitude of sound speed in our experiments were quite small; this is due to the low concentration of the MBs in our experiments, as well as the small size of the microbubbles. Bigger microbubbles have stronger effects on the sound speed changes of the medium because of their higher compressibility as well as their more rapid changes in resonance frequency with increasing acoustic pressure \cite{36}. In many applications (e.g. drug delivery, ultrasound imaging) much higher concentrations of microbubbles are employed, thus greater changes in the sound speed amplitude is expected (e.g. please refer to figure 1d with $\beta =10^{-5}$). Nevertheless, although the changes of the sound speed amplitude are small, we could measure thee changes in experiments consistent with model predictions. \\When fitting the shell parameters (\cite{22,43,44,45}), sound speed values are of great help for more accurate characterizations of the shell parameters, especially in case of microbubbles with more complex rheology \cite{47}. There may exist multiple combination of values of the initial surface tension, shell elasticity and shell viscosity that fits well with measured attenuation curves; however, only one combination provides good fit to both the measured sound speed and attenuation values. Sound speed curves provide more accurate information on the effect of shell parameters on the bulk modulus of the medium (e.g. Shell elasticity) while attenuation graphs are more affected by damping parameters; thus the attenuation and sound speed curves can be used in parallel in order to achieve a more accurate characterization of the shell parameters.\\
In summary, we have presented a nonlinear model for the calculation of the
pressure-dependent attenuation and sound speed in a bubbly medium. The model is
free from any linearization in the MB dynamics. The accuracy of the model
was verified by comparing it to the linear model \cite{17} at low pressures and the
semilinear Lousinard model \cite{18} at higher-pressure amplitudes. The relationship
between the sound speed and pressure was established both theoretically and
verified experimentally. The results of the model predictions are in good
agreement with experimental observations. To our best knowledge,
unlike current sound speed models, the model introduced in this paper does not have a $\frac{dP}{dV}$ term (e.g.  \cite{28}); thus it does not encounter difficulties addressing the
nonlinear oscillations. To accurately model the changes of the attenuation and
sound propagation in a bubbly medium we need to take into account how the
sound speed changes with pressure and frequency; otherwise, prediction accuracy
decreases, especially for US exposure parameter ranges where the sound speed
undergoes larger deviations compared to linear predictions. Another advantage of
the model is that it uses as input only the radial oscillations of the MBs. There
is no need to calculate the energy loss terms, and thus our approach is simpler
and faster.  Moreover, for the case of nonlinear shell behavior (e.g. \cite{16,46}) it may provide
more accurate estimates since there is no need for simplified analytical
expressions. MB oscillator exhibits stable nonlinear oscillations \cite{36,46,47,48,49,50,51,52,53,54} (e.g. period 3, 4 or super and ultra harmonics) and the effect of these nonlinear oscillations on the changes of sound speed and attenuation is not understood. Application of this model will help to shed light on the effect of nonlinear oscillations on the acoustical properties of the bubbly medium (we first reported on this in \cite{31}) and explore new potential parameters to further optimize and improve the current applications.\\
\textbf{Acknowledgments}\\
The work is supported by the Natural Sciences and Engineering Research Council of Canada (Discovery Grant RGPIN-2017-06496), NSERC and the Canadian Institutes of Health Research ( Collaborative Health Research Projects ) and the Terry Fox New Frontiers Program Project Grant in Ultrasound and MRI for Cancer Therapy (project $\#$1034). A. J. Sojahrood is supported by a CIHR Vanier Scholarship and Qian Li is supported by NSF CBET grant $\#$1134420.

\end{document}